\documentclass{aa}  
\usepackage{graphicx}
\usepackage{natbib}
\bibpunct{(}{)}{;}{a}{}{,} 
\begin{document}
\title{ Two different evolutionary types of comets proved by polarimetric and infrared properties of their dust}

   \author{L. Kolokolova \inst{1} \and H. Kimura
 \inst{2} \and  N. Kiselev \inst{3,4} \and V. Rosenbush \inst{3}}
   \offprints{L. Kolokolova\\ email:ludmilla@astro.umd.edu}
   \institute{Dept. of Astronomy, University of Maryland, USA 
\and Institute of Low Temperature Science, Hokkaido University, Japan
\and Main Astronomical Observatory, NAS of Ukraine, Ukraine                                  
\and Institute of Astronomy, Kharkiv National University, Ukraine}


   \date{Received February 22, 2006; accepted        }

 
\abstract
{}
{We consider polarimetric and thermal-emission properties of comet dust and show how and why they can be used for classification of comets.}
{We provide a statistical analysis of comet polarimetric, thermal, and orbital characteristics. We perform computer simulations of  polarization and infrared spectra considering comet particles as ballistic particle-cluster and cluster-cluster aggregates (BPCA and BCCA) consisting of submicron spherical grains.}
{Comets can be divided into two groups: Type I, characterized by high gas/dust ratio, low polarization, and a weak or absent $10~\mu$m silicate feature, and Type II, for which a low gas/dust ratio, high polarization, and strong silicate feature are typical. We show that the low polarization is the apparent result of depolarization by gas contamination at low dust concentration, which, in turn, results from the dust in Type I comets being concentrated near the nucleus. The simulations of thermal emission show that for more porous particles (BCCA), the silicate feature is more pronounced than  more compact ones (BPCA), for which it even vanishes as the particles become larger. We also show that in both types of comets the main contribution to light scattering and emission comes from particles larger than 10 micron.}
{The strength of the silicate feature in the cometary infrared spectra suggests that the dust in Type II comets consists of high-porosity aggregates, whereas the dust of Type I comets contains low-porosity ones. This is consistent with the polarimetric features of these comets, which indicate that the dust in Type I comets tends to concentrate near the nucleus. This may result from the predominance of highly processed particles in Type I comets, whereas in Type II comets we see pristine or slightly-processed dust. This conclusion is in accordance with the orbital characteristics of the comets. We have found that the strength of the silicate feature correlates with the semi-major axis of periodic comets and, for short-period comets, with the perihelion distance. Thus, the silicate feature weakens due to compaction of aggregate particles if a comet spends more time in the vicinity of the Sun, which allows the comet to accumulate a mantle on the surface of its nucleus.}


\maketitle

\section{Introduction}
The idea that properties of cometary dust can be a basis for comet classification appeared after \citet{chernova93} and \citet{LR1996} noticed that comets can be divided into two groups according to the values of the polarization of their dust at phase angles $\alpha \approx 80-100^\circ$: the polarization tends to be larger than 20\% for one group of comets and smaller than 15\% for the other group. \citet{chernova93} also showed that high-polarization comets are characterized by low gas/dust ratios. Quantitatively, they followed the approach of \citet{KrishnaSwamy86} and characterized the dust abundance in comets using the ratio of the fluxes in the C$_2$ band centered at $\lambda=5140$~\AA\, and the nearby continuum at $\lambda = 4845$~\AA.  This ratio is denoted as $W$ and is measured in \AA, since the flux in the C$_2$ band is integrated over the whole band, whereas the flux for the continuum is measured for a unit of wavelength. Parameter $W$ can be estimated directly from photometric or polarimetric observations. It characterizes the gas/dust ratio in the coma at the moment of the observations and uses the same wavelengths as those used to measure the polarization, whereas rigorous determination of the gas/dust ratio, which is usually defined as the mass ratio of water to the dust, requires special infrared observations and has most often appeared to be done for different dates from when polarization was measured.  It turned out that low-polarization comets have high values of $W$, $W>1000$~\AA; they are namely characterized by strong gas emission in comparison with the continuum. This allowed the comets with high $W$ to be called ``gas-rich comets'' whereas the other ``dust-rich comets'' usually have $W<500$~\AA. It was also found \citep{chernova93, LR1996, hanner} that comet polarization correlates with the thermal infrared characteristics: higher polarization is usually accompanied by stronger 10~$\mu$m silicate feature and higher dust temperature. 

In this paper we consider the polarimetric and thermal properties of comet dust. We provide their interpretation based on recent observational data and theoretical simulations of the dust light scattering and emission. We present updated data on polarization for dust-rich and gas-rich comets and show that they indicate a different distribution of the dust within the coma.  Our theoretical simulations are used to analyze which properties of the dust particles influence the polarization and strength of the silicate infrared feature. This, together with the orbital characteristics of the comets, allows us to reveal the reasons for the existence of the two groups (types) of comets.

\section{Polarimetric properties of comet dust: maximum polarization}

The dependence of comet polarization on phase angle is usually described by (1) position and value of the minimum (negative) polarization, (2) position of the inversion angle (the phase angle where polarization changes from negative to positive) and the polarization angular gradient, $dP/d\alpha$, at the inversion point, and (3) position and value of the maximum (positive) polarization. Amazingly, all comets show very similar characteristics of negative polarization (position of the minimum at $\alpha=10^\circ$ with the value $P\approx-1.5$\%, inversion angle at 20--22$^\circ$, polarimetric slope $dP/d\alpha \approx 0.3$\%/deg). However, \citet{chernova93} and \citet{LR1996} found that at $\alpha > 35^\circ$, the polarization curve ``forks'', forming two groups of curves with high and low maximums of polarization. The two groups of comets are seen in the second column of Table~\ref{tab1}. The numbers in this column were estimated from the data presented in \citet{Kiselev-basa}. Very often comets cannot be observed at the phase angles of the maximum polarization, i.e. $\alpha \approx 95^\circ$. Therefore in the majority of cases the numbers in column 2 represent not the measured but the expected values of the maximum polarization, estimated from the data obtained at $\alpha > 35^\circ$.  The least reliable data are for comets 9P/Tempel 1 and 22P/Kopff, which were observed at the phase angles whose maximum values reached $41^\circ$ and $37^\circ$ correspondingly. However, at these phase angles the difference between dust-rich and gas-rich comets can already be seen. Both comets showed polarization that exceeded the values typical of these gas-rich comets (approximately 2-3\% higher), and the observed tendency was close to the one typical of the dust comets. 
The table lists only those comets whose infrared characteristics are also known as we need this for the following analysis (see Sect. 3).  The third column of Table~\ref{tab1} summarizes the data on the comet gas/dust ratio as described in the Introduction. The fourth column shows the references from which the values of parameter $W$ or other characteristics of the gas/dust ratio were taken. Table~\ref{tab1} shows that high polarization is typical of comets with low values of $W$, i.e. with a low gas/dust ratio. 
\begin{table*}
\caption{Comet gas/dust characteristics and polarization.}           
\label{tab1}                               
\begin{minipage}[t]{18cm}
\centering
\renewcommand{\footnoterule}{}  
\begin{tabular}{l c c c c }        
\hline\hline                
Comet name &Polarization\footnote{the data for the red spectral range} &W&  Reference on \\   
& at $\alpha \approx 95^\circ$&\AA &gas/dust ratio \\
\hline                      
2P/Encke&$\sim 8-10\%$\footnote{as observed through wide-band filters and large diaphragm \citep{Jockers2005}}&weak continuum&\citet{Ahearn}\\
C/1975 N1 (Kobayashi-Berger-Milon)&$\approx 10\%$ &weak continuum&\citet{Winiarskietal92}\\
23P/Brorsen-Metcalf&10--15\% &1840-2010&\citet{chernova93}\\
27P/Crommelin&10-15\%& 860-1470&\citet{chernova93}\\
C/1989 Q1 (Okazaki-Levy-Rudenko)&$\approx 16\%$&gas-rich comet&\citet{Winiarskietal92}\\
D/1996 Q1 (Tabur)&12—-17\%\footnote{at large diaphragms \citep{Jockers97}} &gas-rich comet&\citet{Kawakita}\\
C/1989 X1 (Austin)&$< 20\%$&180-1300&\citet{Joshi}\\
67P/Churyumov-Gerasimenko& $\leq 20\%$&relatively dust-rich comet&\citet{Hanner-Churyumov}\\
9P/Tempel 1&$\approx 24\%$&dust-rich comet&\citet{kuppers-et-al2005}\\
21P/Giacobini-Zinner&$\approx 24\%$&50&\citet{chernova93}\\
C/1983 H1 (IRAS-Araki-Alcock)& $>$25\%\footnote{at small diaphragm}&dust-poor comet&\citet{Hanner85}\\
22P/Kopff&$\approx 26\%$&30&\citet{chernova93}\\\
1P/Halley&$\approx 26\%$&100&\citet{chernova93}\\
C/1987 P1 (Bradfield)&$\approx 26\%$&160&\citet{chernova93}\\
C/1990 K1 (Levy)&$\approx 26\%$&220&\citet{chernova93}\\
C/1996 B2 (Hyakutake)&$\approx 26\%$&200&\citet{Velichko}\\
C/1975 V1 (West)&$\approx 26\%$&strong continuum&\citet{Rosenbush}\\
C/1995 O1 (Hale-Bopp)&$> 30\%$&120-160&\citet{Velichko2}\\
\hline                                 
\end{tabular}
\end{minipage}
\end{table*}
In the following part of this section we explore possible reasons for the existence of two groups of comets that are different in their polarization and gas/dust ratio. 

We start with analyzing our computer simulations of light scattering by cometary dust. Using the T-matrix technique by \citet{mackowski-mishchenko96}, we calculated light scattering by the most realistic type of cometary grains \citep[see][]{comets2}: aggregates of submicron particles. We considered random aggregates built using the ballistic procedure described in \citet{BPCA,BCCA}. Two types of aggregates, more compact ballistic particle-cluster aggregates (BPCA) and more porous ballistic cluster-cluster aggregates (BCCA) were considered.  For both types of aggregates we managed to obtain the correct behavior for the angular and spectral dependences of comet brightness and polarization, as well as the correct change in spectral characteristics (color, polarimetric color) with the phase angle \citep{Kimura2003}. We used the Halley-type composition of the dust material and the constituent particles (monomers) with a radius of $0.1~\mu$m.  More detailed calculations and their analysis are presented in \citet{kimura-et-al06}, where we analyzed how the size of monomers and their composition influence the angular and spectral characteristics of the scattered light. Our calculations show \citep[similar trends can be also noticed in the results obtained by] [] {Lasue} that the maximum polarization gets smaller (keeping the correct shape of the curve $P(\alpha$)) under the following conditions:
\begin{itemize}
\item (1) if the real part of the refractive index gets larger;
\item (2) if the imaginary part of the refractive index gets smaller in the range within 0.2--0.8, while the real part of the refractive index is larger than 1.8; 
\item (3) if the monomers in the aggregates get larger.
\end{itemize}
Could any of the reasons listed above reasons be responsible for the difference in the maximum polarization in two groups of comets? The calculations by \citet{kimura-et-al06} show that the reasons for changing the maximum polarization also influence other characteristics of the scattered light making them unrealistic. Thus, reasons (1) and (2) change the color of the scattered light to blue, whereas the color of the comet dust is usually red \citep[blue colors observed for some gas-rich comets are artifacts from gas contamination, see] []{Ahearn}. Reason (3) leads to some incorrect behavior by the negative polarization. We conclude that a difference in comet dust size or composition could not explain the difference in comet maximum polarization without producing a contradiction with the rest of the observational facts.  Our numerical simulations instead suggest that polarization should be intrinsically high regardless of the properties of the dust particles.

\begin{figure}
\centering
\includegraphics[angle=270, width=6.5cm]{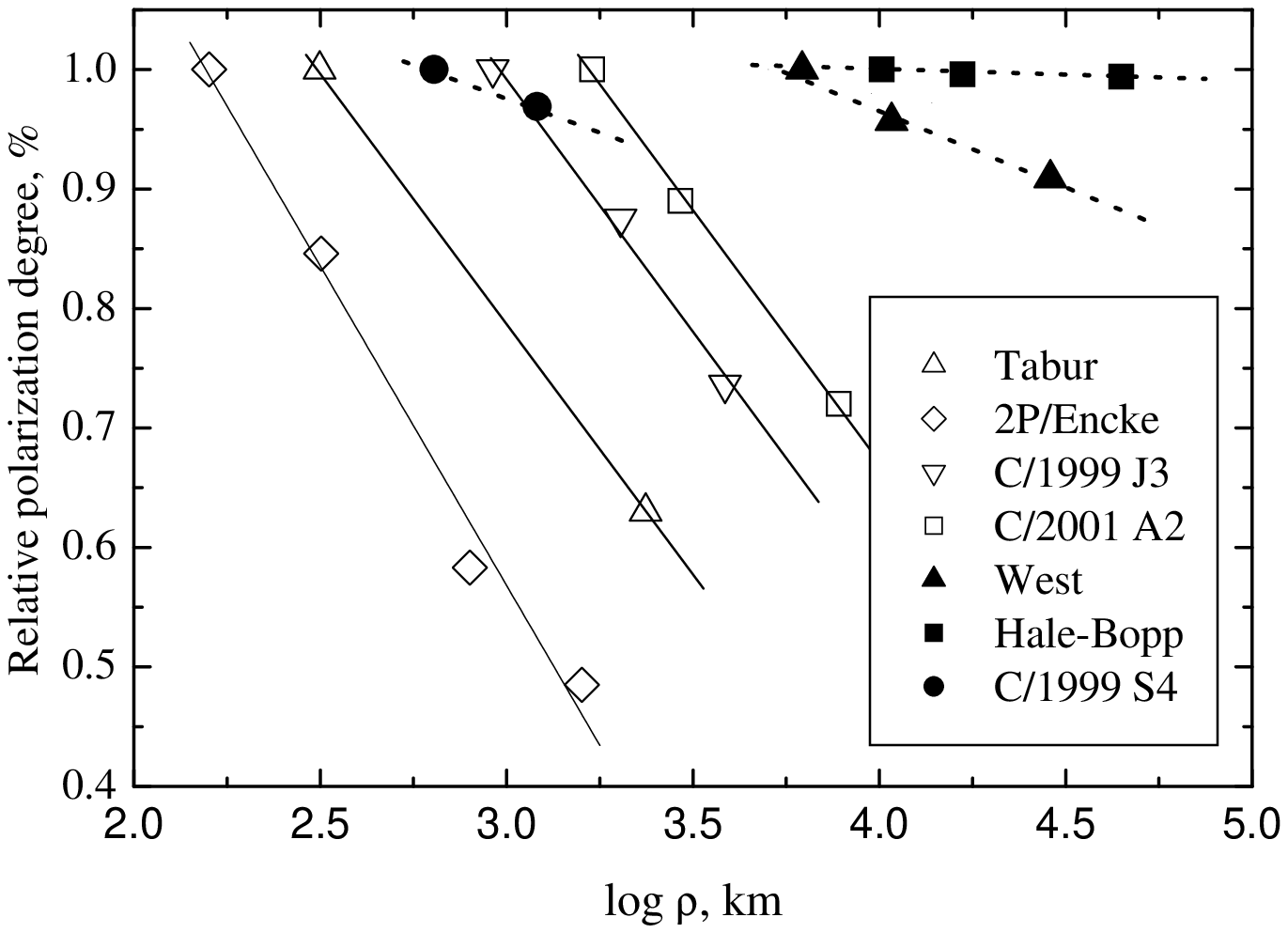} 
\caption{Comet polarization in the red spectral range versus the distance from the nucleus for dust-rich (filled symbols) and gas-rich (open symbols) comets. The polarization is averaged over the aperture of radius $\rho$ and is normalized to the value obtained at the smallest aperture.}
\label{fig1}
\end{figure}

\begin{figure}
\includegraphics[width=9.0cm]{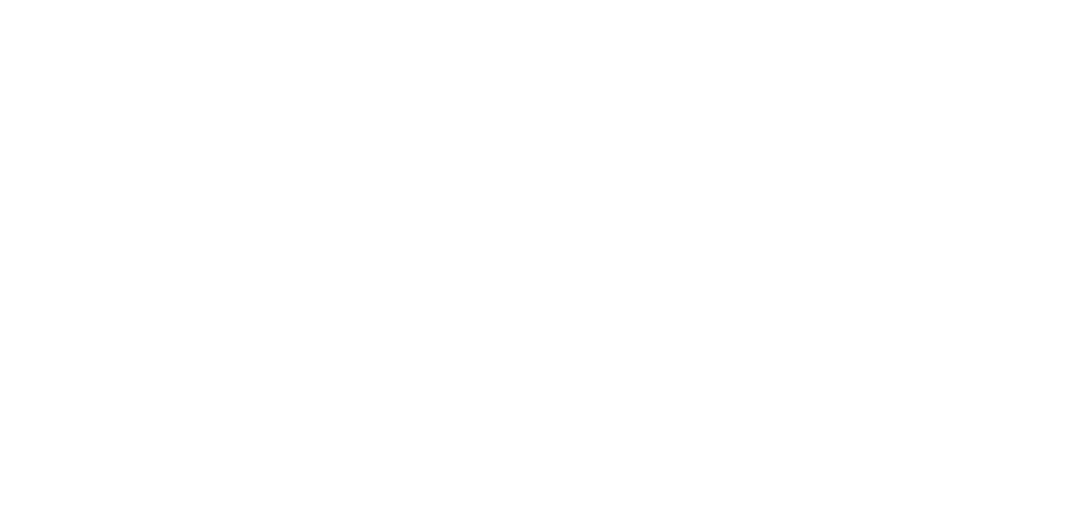} 
\caption{Phase-angle dependence of polarization for dust-rich (left panel) and gas-rich(right panel) comets in the red spectral range \citep[based on][] {Kiselev-basa}. The solid line shows the best fit curve for dust-rich comets. At large phase angles, polarization averaged over a large aperture for gas-rich comets (large symbols) is substantially smaller than for dust-rich comets illustrating the original idea about two groups of comets. Polarization of gas-rich comets measured in the near-nucleus area of the coma and properly reduced for gas contamination is significantly larger (small symbols). Vertical lines show the range of polarization change as the distance from the nucleus (aperture) changes.} 
\label{fig2}
\end{figure}

Recent observational data \citep{Kiselevetal2001, Kiselev2004, Jewitt, Jockers2005} have shed new light on the two groups of polarimetrically different comets. They show that low polarization manifests itself either when the polarization is averaged over a large area in the coma or far from the nucleus. Approaching the nucleus, the polarization increases and, in the near-nucleus region, reaches the values typical of high-polarization comets. The difference in the behavior of polarization with the distance from the nucleus for comets with different polarimetric properties is seen in Fig. ~\ref{fig1}.  
As Table~\ref{tab1} demonstrates, comets with low polarization are also characterized by the predominance of gas over dust in the coma. Based on this, \citet{Kiselevetal2001} and then \citet{Jockers2005} explain the low average polarization by the contribution of gas emission in the continuum, which causes the depolarization of the continuum light. Their analysis was based on a detailed comparison of the wavebands of narrow-band filters and corresponding parts of almost-simultaneously-obtained cometary spectra, and it allowed direct determination of gas contribution into continuum. They find that as we get closer to the nucleus, concentration of the dust increases. This decreases the relative contamination by gas emission and makes the polarization values similar to those observed for the comets with dust-dominated comae. The coma of gas-rich comets gets dominated by gas at nucleocentric distances $> 1000$~km; this is why at the observations with low spatial resolution \citep[see, e.g.][]{Kelley} the spatial gradient in polarization reported by \citet{Jockers2005} and \citet{Jewitt} may not be noticed. If we make a plot that includes the data for different distances from the nucleus, the comets still can be divided in two groups. However, the difference between the groups will be in the different distributions of the polarization over the coma. In dust-rich comets, polarization does not show significant changes with the distance from the nucleus. Thus, dust-rich comets can be characterized by some typical value of maximum polarization (Fig.~\ref{fig2}, left panel). Although some difference between comets can be noticed, e.g. higher polarization for Hale-Bopp or lower polarization for Giacobini-Zinner, this difference is a property of the comet itself (and may indicate some sub-classes of comets), independent of the aperture or filter passband. In the case of gas-rich comets, the polarization depends significantly on the aperture of the measurements and filter passband (Fig. ~\ref{fig2}, right panel), thus these comets cannot be characterized by a single ``typical'' value of the maximum polarization.

Morphological features such as jets and shells can show somewhat different polarization from the polarization of the rest of the coma. However, their polarization does not contribute significantly to the integral polarization of comets, especially at large phase angles.  This can be seen from the fact that the dependences of polarization vs. phase angle for dust-rich comets are stable and do not demonstrate any noticeable scattering of the data. The more noticeable scattering of the data for gas-rich comets can be explained by the different contribution of gas for different comets  \citep[see, ][]{Kiselev2004, Jockers2005}.
In summary, the difference in the comet polarization results not from the different size or composition of the dust grains, but from the fact that in low-polarization comets the dust is concentrated near the nucleus, leaving distant parts of the coma dominated by gas, which contaminates the continuum and, thus, depolarizes it. However, this explanation does not answer why comets form two groups. It just shifts the question about two polarimetric groups of comets to another one: why are there two groups of comets that differ in the distribution of the dust in the coma? We try to find the answer in the following section using computer simulations of the thermal properties of cometary dust.

\section{Infrared properties of comet dust: strength of the silicate feature}

As mentioned in the Introduction, the values of maximum polarization correlate with some thermal characteristics of comets, e.g. dust temperature and the strength of the 10~$\mu$m silicate feature. Since there is a strong correlation between the temperature and strength of the silicate feature \citep[see, e.g.,][] {GN1992, Lisse2003}, we will characterize comet thermal emission only by the strength of the silicate feature, considering the dust temperature as a redundant characteristic. Following \citet{Lisse2002} and \citet{Sitko}, we define the strength of the silicate feature as the ratio of the flux between 10 and 11 micron to that of the underlying continuum. Although the infrared feature of crystalline silicate differs from that of amorphous silicate, this does not influence the strength of the silicate feature; a detailed analysis of the silicate mineralogy is the subject of our separate study. 

Table \ref{tab2} summarizes the data on the strength of the silicate feature and demonstrates that comets can be divided into two groups, which we call Type I (the comets with a weak silicate feature) and Type II (the comets with a strong silicate feature) comets following \citet{GN1992}. The first idea that comes to the mind is that the strong silicate feature indicates high abundance of silicates in the dust, whereas the low silicate feature shows their shortage. However, comparing the second columns in Table~\ref{tab1} and Table~\ref{tab2}, one can see a correlation between the values of maximum polarization and the strength of the silicate feature: Type I comets are also characterized by low polarization whereas Type II comets have high polarization.  In Sect. 2 we showed that polarimetric properties do not indicate any significant compositional difference between the dust in different comets.
 
\begin{table*}
\caption{Infrared and orbital characteristics of comets.}
\label{tab2}     
\begin{minipage}[t]{18cm}
\renewcommand{\footnoterule}{} 
\centering                          
\begin{tabular}{l c c c c c}        
\hline\hline                
Comet name &Silicate &Perihelion, &Eccentricity & Semi-major&Reference\\   
&feature strength\footnote{Although no distinct dependence of the strength of the silicate feature on the heliocentric distance \citep{Harker-etal, Sitko}was observed, in the case where several values were available we selected the value at the heliocentric distance $\sim$ 1 AU. Note also that \citet{hanner2005} find no difference between pre- and post-perihelion silicate features.} &AU&&axis, AU&for infrared data \\
\hline                      
2P/Encke&$<$1.1&0.33&0.85&2.2&\citet{Lisse2002}\\
&&&&&\citet{Gehrz}\\
C/1975 N1 (Kobayashi-Berger-Milon)&~1.0&0.425&1.00097&$\infty$&\citet{Ney}\\
23P/Brorsen-Metcalf&~1.0&0.478&0.97&15.9&\citet{Lynch-etal}\\
27P/Crommelin&1.1&0.743&0.92&9.29&\citet{EZ}\\
C/1989 Q1 (Okazaki-Levy-Rudenko) & 1.2& 0.642 & 1.0000197 & $\infty$& \citet{Sitko}\\
D/1996 Q1 (Tabur)&1.3&0.842&0.999474&1600&\citet{Harker-etal}\\
C/1989 X1 (Austin)&1.13&0.349&1.000225&$\infty$&\citet{Sitko}\\
67P/Churyumov-Gerasimenko&1.0&1.292&0.63&3.49&\citet{Hanner-Churyumov}\\
9P/Tempel 1&$<$1.1&1.5&0.52&3.13&\citet{Lisse2002}\\

C21P/Giacobini-Zinner&1.1&1.034&0.71&3.56&\citet{Hanner-GZ}\\
C/1983 H1 (IRAS-Araki-Alcock)&1.04&0.991&0.990115&100.2&\citet{Sitko}\\
22P/Kopff&1.2&1.58&0.54&3.43&\citet{Lisse2002}\\
1P/Halley&1.6&0.587&0.97&19.6&\citet{Harker-etal}\\
C/1987 P1 (Bradfield)&1.9&0.869&0.999697&2868&\citet{Harker-etal}\\
C/1990 K1 (Levy)&1.8&0.939&1.000417&$\infty$&\citet{Harker-etal}\\
C/1996 B2 (Hyakutake)&$>$1.5&0.230&0.999758&950.4&\citet{Lisse2002}\\
C/1975 VI (West)&2.0&0.196&0.999971&6752&\citet{Ney}\\
C/1995 O1 (Hale-Bopp)&2.16&0.914&0.995069&185.4&\citet{Sitko}\\
\hline
\multicolumn{6}{c}{Coefficient of correlation between strength of the silicate feature and corresponding orbital characteristic}\\                                
\hline
All comets&& -0.193\\
Periodic comets&& -0.352&&0.662&\\
Short-period comets &&0.718&&0.751&\\
\hline
\end{tabular}
\end{minipage}
\end{table*}
To find the reason for the difference in the infrared properties of the comet dust, we calculated infrared spectra of aggregated particles choosing their size, composition, and structure the same as provided the best fit to the observed brightness and polarization in the calculations presented in \citet{kimura-et-al06}, which were discussed in Sect. 2. We built our particles as ballistic aggregates and, to account for different porosity, considered BCCA and BPCA particles. Calculations were performed for an aggregate made of spherical monomers of radius 0.1 $\mu m$ whose composition corresponds to the Halley-comet material, i.e. we assume that cometary dust is composed of magnesium-rich olivine, organic refractory, amorphous carbon, and pyrrhotite with their volume fractions derived from \citet{kimura-et-al06}. We took the refractive indices for magnesium-rich olivine from \citet{mukai-koike90}, organic refractory from \citet{li-greenberg97}, amorphous carbon from \citet{rouleau-martin91}, and pyrrhotite from \citet{begemann-et-al94}, with the refractive indices of pyrrhotite below $\lambda=10~\mu$m extrapolated from the near-infrared data of \citet{egan-hilgeman77}. Table~\ref{tbl-1} presents the effective refractive indices of this synthetic mixture as a function of wavelength calculated using the Maxwell Garnett mixing rule \citep{bohren-huffman83}.
\begin{table}
\caption{Complex refractive indices ($m=n+\mathrm{i}\,k$) at a wavelength $\lambda$ for the synthetic materials.}             
\label{tbl-1}      
\centering                          
\begin{tabular}{l l}        
\hline\hline                 
Wavelength      & Complex refractive index \\    
$[\mu\mathrm{m}]$ & $m=n+\mathrm{i}\,k$ \\
\hline                        
8.0 & $2.67+\mathrm{i}\,6.11\times{10}^{-1}$\\      
8.5 & $2.70+\mathrm{i}\,6.23\times{10}^{-1}$\\
9.0 & $2.71+\mathrm{i}\,6.33\times{10}^{-1}$\\
9.5 & $2.70+\mathrm{i}\,6.58\times{10}^{-1}$\\
10.0 & $2.71+\mathrm{i}\,7.82\times{10}^{-1}$\\
10.5 & $2.84+\mathrm{i}\,7.12\times{10}^{-1}$\\
11.0 & $2.72+\mathrm{i}\,7.61\times{10}^{-1}$\\
11.5 & $2.94+\mathrm{i}\,1.16\times{10}^{-1}$\\
12.0 & $3.16+\mathrm{i}\,7.71\times{10}^{-1}$\\
12.5 & $3.14+\mathrm{i}\,6.07\times{10}^{-1}$\\
13.0 & $3.03+\mathrm{i}\,6.08\times{10}^{-1}$\\
\hline                                   
\end{tabular}
\end{table}
We estimate the absorption cross sections for model aggregates in the infrared wavelengths using Mie theory along with the Maxwell Garnett mixing rule \citep[see][]{mukai-et-al92}. Thus, we calculate thermal emission for a spherical particle with the radius chosen to provide a characteristic size of the corresponding aggregate. The refractive index of the sphere material is calculated using the Maxwell Garnett mixing rule \citep{bohren-huffman83} when the material was presented as a mixture of the aggregate material from Table~\ref{tbl-1} and the voids, so that the ratio of the volume occupied by the material to the volume of the voids corresponds to those in the aggregate. 

Our numerous tests showed that the spectral dependence of the thermal emission calculated with the Maxwell Garnett approximation keeps all the features of the spectral dependence obtained with more rigorous methods. At the number of monomers $N=2^{10}$, an error in the infrared absorption cross section arising from this approximation is evaluated using the superposition T-matrix method, which gives a rigorous solution \citep{mackowski-mishchenko96}. For an alternative error estimate, we also calculate the infrared absorption cross sections for aggregates consisting of $N=2^{15}$ monomers using the discrete dipole approximation \citep{okamoto-et-al94}. We find that errors are nearly independent of wavelength, and thus our qualitative discussion on the infrared spectra can be validated.
\begin{figure*}
\centering
\includegraphics[width=18cm]{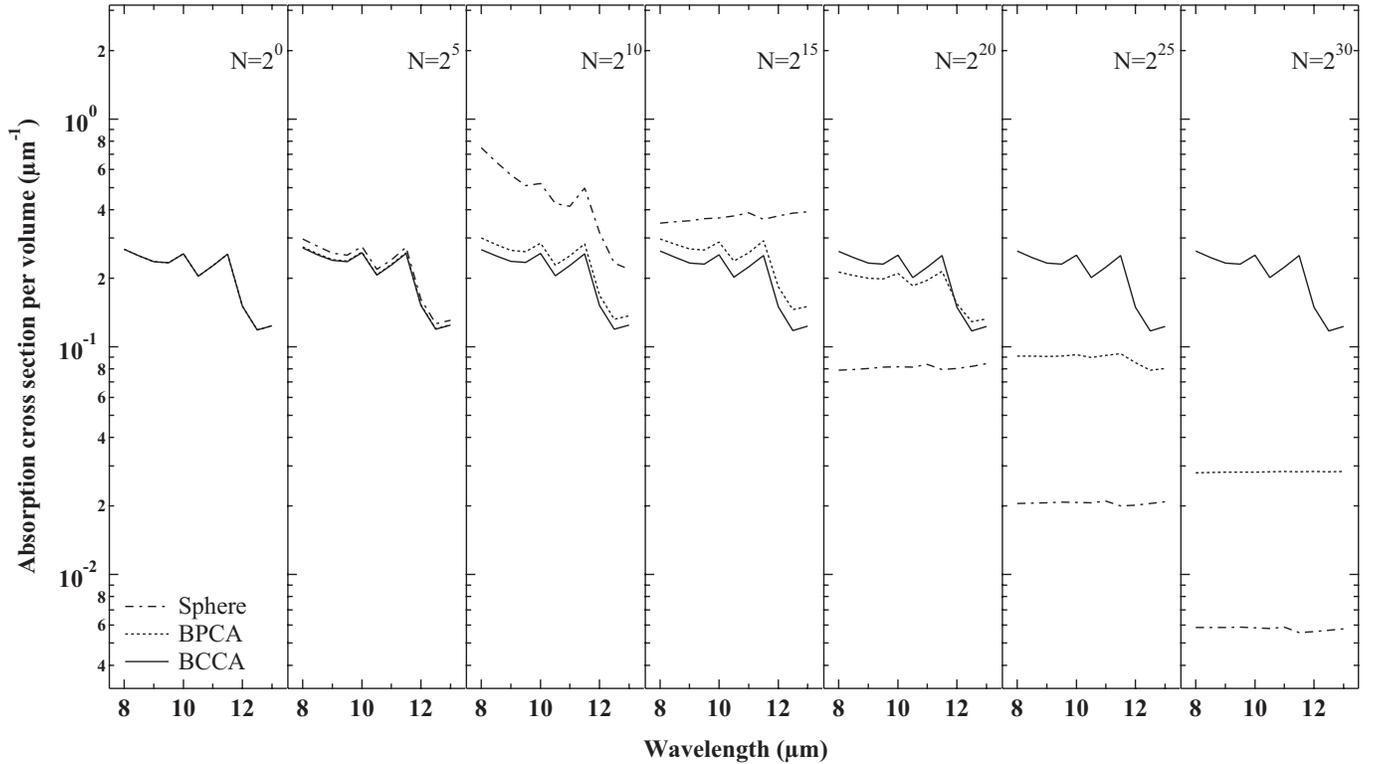} 
\caption{Absorption cross section per volume as a function of wavelength for aggregate particles composed of a synthetic mixture of magnesium-rich olivine, amorphous carbon, and pyrrhotite. The number of monomers increases from $N=2^0$ (left) to $2^{30}$ (right), which correspond to the radii of aggregates ${a}_{\mathrm{V}}=0.1$--$100~\mu$m. Solid lines: results for particles grown under the ballistic cluster-cluster aggregation (BCCA) process; dotted lines: results for particles grown under the ballistic particle-cluster aggregation (BPCA) process; dash-dotted lines: results for compact homogeneous spherical particles.}
\label{fig3}
\end{figure*}

\begin{table}
\caption{The porosity P of BPCA and BCCA at various sizes.}
\label{tbl-2}
\begin{minipage}[t]{\columnwidth}
\centering
\renewcommand{\footnoterule}{}  
\begin{tabular}{llcllcll}
\hline \hline
& ${a}_{\mathrm{V}}$\footnote{The radius of volume-equivalent spheres.} &&\multicolumn{2}{c}{$a_\mathrm{C}$\footnote{The characteristic radius of aggregate particles: $a_\mathrm{C} = \sqrt{5/3}\,a_\mathrm{g}$
where $a_\mathrm{g}$ denotes the gyration radius of the aggregates.}
[$\mu$m]} && \multicolumn{2}{c}{$1-P$\footnote{$1-P=N\,(a/a_\mathrm{C})^3$ where $a$ is the radius of monomers.}}
 \\
  \cline{4-5}\cline{7-8}
$N$  & [$\mu$m] && BPCA & BCCA && BPCA & BCCA \\
\hline
$2^{0}$ &$0.10$&&$0.10$&$0.10$&&$1.0$ & $1.0$\\
$2^{5}$ &$0.32$&&$0.59$&$0.76$&&$0.15$& $0.072$\\
$2^{10}$&$1.0$&&$1.9$&$4.6$&&$0.15$& $0.011$\\
$2^{15}$&$3.2$&&$6.1$&$28.0$&&$0.15$& $1.6\times{10}^{-3}$\\
$2^{20}$&$10.0$&&$19.0$&$170$&&$0.14$& $2.3\times{10}^{-4}$\\
$2^{25}$&$32.0$&&$62.0$&$1000$&&$0.14$& $3.3\times{10}^{-5}$\\
$2^{30}$&$100$&&$200$&$6000$&&$0.14$& $4.9\times{10}^{-6}$\\
\hline
\end{tabular}
\end{minipage}
\end{table}
Figure~\ref{fig3} presents the absorption cross sections divided by the volume of aggregate particles composed of the synthetic mixture of magnesium-rich olivine, organic refractory, amorphous carbon, and pyrrhotite. Each panel shows the results for aggregates of different sizes, defined by the number of monomers $N$. One can see that the infrared spectra for those aggregates smaller than $N \le 2^{20}$ (${a}_{\mathrm{V}} \le 10~\mu$m where $a_{\rm V}=a N^{1/3}$) always show the silicate features irrespective of their sizes and porosities. When the aggregates reach radii ${a}_{\mathrm{V}} > 10~\mu$m, low-porosity (Table~\ref{tbl-2}) BPCA particles do not show the silicate features, but high-porosity (Table~\ref{tbl-2}) BCCA particles retain the features. 

In the previous studies \citep[see, e.g.,][]{hanner, Lisse2002,Lisse2003, Lisse2004}, weak or absent silicate features and corresponding low polarization were explained by the predominance of large particles in the coma, whereas strong silicate features and high polarization were presumed, indicating small particles. This conclusion might be correct if the low polarization in comets with a weak silicate feature were an intrinsic property of the dust. However, as we showed in Sect. 2, all comets in the near-nucleus area have similar (high) values of polarization. This could be impossible if the particles have different sizes \citep[see Sect. 2 and] []{Kimura2003, kimura-et-al06}. As we concluded earlier, two types of comets are characterized by different distributions of the dust in the coma: the dust tends to concentrate near the nucleus for low-polarization comets. 

Thus, we have to look for another characteristic than size of the dust particles that can be responsible for the different strength of the silicate feature. The analysis of Fig.~\ref{fig3} allows us to suggest that the porosity of the dust particles may be the very crucial parameter that divides comets into two groups.
Dust particles are porous in the comets that demonstrate a strong silicate feature and more compact in the comets with a weak or absent silicate feature. This conclusion is consistent with the different spatial distribution of the dust in the comets of different types. Indeed, compact large particles in the ``low-polarization'' comets are heavy and cannot be accelerated by the gas drag, whereas porous, albeit large, particles in ``high-polarization'' comets can easily be moved out of the nucleus.

\section{Physical properties of comet dust: typical sizes}

\begin{figure}
\centering
\includegraphics[width=7cm]{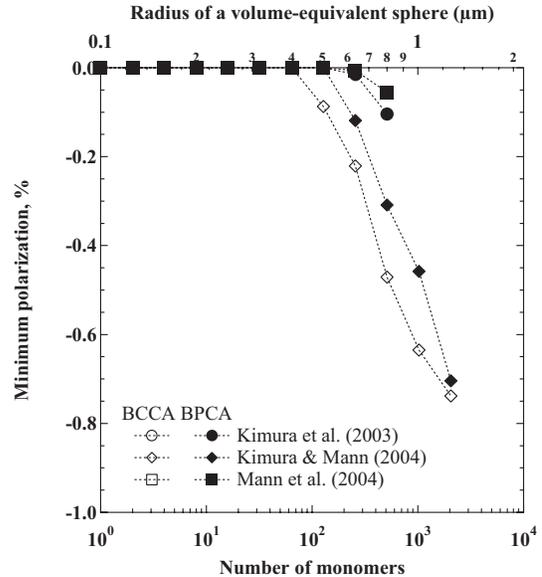} 
\caption{Computed minimum polarization as a function of the size of aggregate.}
\label{fig4}
\end{figure}

\begin{figure}
\centering
\includegraphics[width=7cm]{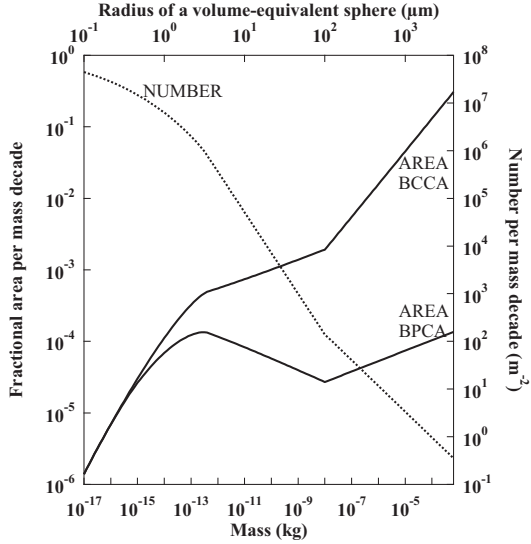} 
\caption{Size distribution from the \textit{in situ} measurements for comet Halley \citep{McDonnell} and corresponding cross-section area of the dust particles for BPCA and BCCA.}
\label{fig5}
\end{figure}

Before reasoning out the difference in dust porosities, we look into a typical size of aggregates.
We have noticed from our numerical calculations of linear polarization that large aggregates are required to produce the observed negative polarization at small phase angles \citep{kimura-mann2004,Kimura2003,kimura-et-al06}.
In Fig.~\ref{fig4}, we summarize our previous results for the dependence of the minimum polarization on the size of aggregates at a wavelength of 0.6~$\mu$m, including our unpublished results.
This clearly shows that the observed negative polarization of approximately $-1.5$\% cannot be attained by small aggregates of a few microns or less.
To be consistent with both polarimetric and infrared observations of comets, we have to admit that the dust particles, which provide the major contribution in the light-scattering and thermal-emission characteristics of comets, are larger than ${a}_{\mathrm{V}}=10~\mu$m for any comet.
This conclusion is also consistent with the \textit{in situ} data for the mass distribution of the dust in comet Halley \citep{McDonnell}. Comet Halley is a typical high-polarization, strong-silicate-feature comet, i.e., according to \citet{Lisse2002}, its dust should be characterized by predominantly small particles. However, the size distribution of cometary dust derived from in situ data on comet 1P/Halley shows that large particles contribute the majority of the total cross-section area \citep{McDonnell, fulle}.  We updated the distribution of the cross-section area of the particles (the parameter, which defines the effectiveness of the light scattering and emission) shown in \citet{McDonnell} considering comet particles as BCCA and BPCA with the density of the monomer material equal to $2.4\times {10^3}$~kg~m$^{-3}$ (based on the composition described in the beginning of Sect. 3). The results are shown in Fig.~ \ref{fig5}.
One can see that the cross-section distribution for BPCA has a local maximum around $\approx 10^{-13}$ kg (${a}_{\mathrm{V}}\approx 3~\mu$m), but reveals the importance of much larger particles having radii of ${a}_{\mathrm{V}}> 10^{3}~\mu$m.
For BCCA particles, the cross section is smoothly increasing with the size of particles, indicating that large particles make a much more significant contribution into light scattering and emission.
\section {Synthesis of the observational data and orbital characteristics of comets}

In this section we try to find out why the dust in some comets is dominated by compact particles, while the particles are porous in other comets. Following the way suggested by numerous previous studies, we check if this comes from the in different evolution of the comets, i.e. if there are any dynamical characteristics of the comets that correlate with the properties of their dust.
\citet{Ahearn} noticed that the perihelion distance correlates with the gas/dust ratio. 
According to \citet{Lisse2002} and \citet{Lisse2003}, there is a correlation between the value of perihelion distance and strength of the silicate feature.
This is why perihelion distance (third column in Table~\ref{tab2}) is the orbital characteristic whose correlation with the strength of the silicate feature we are checking first. Table ~\ref{tab2} also shows eccentricity, $e$, of the considered comets. This allows us to distinguish between new ($e>1$), long-period ($0.99<e<1$), Halley-type ($0.99<e<0.9$), and short-period ($e<0.9$) comets.

\begin{figure}
\centering
\includegraphics[width=7cm]{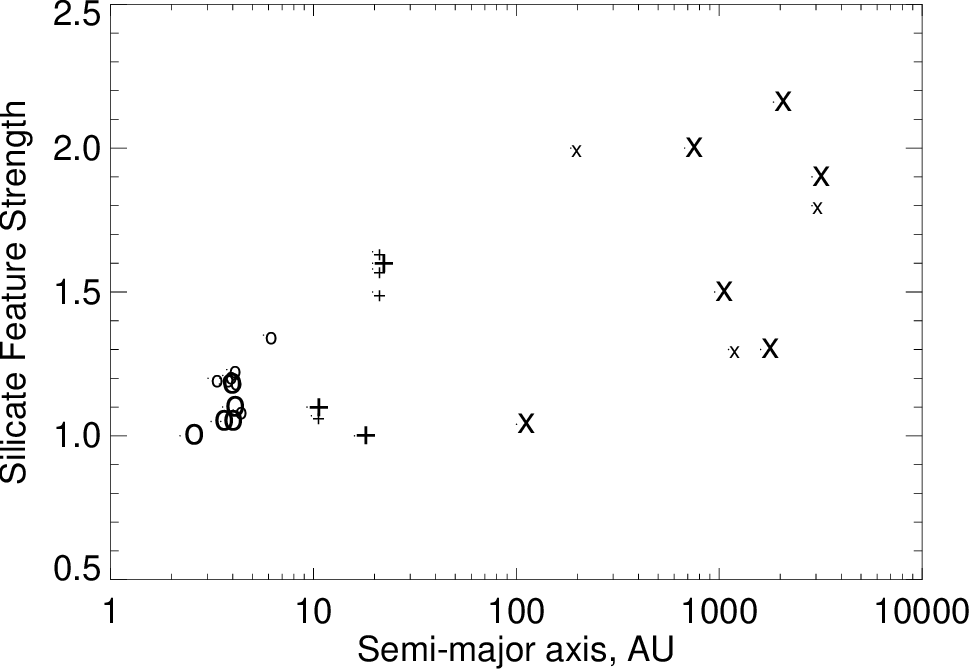} 
\caption{Dependence of the strength of the silicate feature on the semi-major axis of the comet orbit for Jupiter family (o), Halley-type (+), and long-period (x) comets. The comets from Table 2 are shown by larger symbols.}
\label{f6}
\end{figure}

The bottom of Table~\ref{tab2} shows the coefficient of correlation between infrared and orbital parameters. One can see that the correlation between perihelion and strength of the silicate feature is rather significant for short-period comets but loses its significance if long-period comets are included in the analysis. Based on this we suppose that the closeness of a comet to the Sun is a key issue; thus, not the perihelion, but the comet average distance from the Sun (or the semi-major axis of its orbit) may be the characteristic to consider. The fifth column of Table~\ref{tab2} shows the values of this characteristic and its correlation with the strength of the silicate feature. This is also illustrated by Fig.~\ref{f6}. We can see that all periodic comets show a significant correlation with this characteristic. For short-period comets the correlation with the semi-major axis is even more significant than the correlation with the perihelion distance. 

The results of the performed statistical analysis, together with the conclusions of Sect. 3, are summarized in Table~\ref{tab5}. They   allow us to suggest that the more time a comet spends in the vicinity of the Sun and the closer it approaches the Sun, the more compact are its dust particles. 
\begin{table}
\caption{Classification of comets based on their polarimetric, infrared, and orbital characteristics}
\label{tab5}     
\begin{minipage}[t]{\columnwidth}
\renewcommand{\footnoterule}{} 
\centering                          
\begin{tabular}{l l l}        
\hline               
Characteristic &Type I& Type II\\   
\hline                      
Gas/dust ratio             &  high &           low \\ 
Polarization                &  &          \\
 - averaged over  &low& high   \\
~~~~~large aperture & &\\  
 - spatial distribution & increases as  & rather \\
 & approaching & homogeneous \\
 & the nucleus &\\

Silicate feature strength   & low        &    high\\
Average distance  &small       &   large\\
~~from the Sun\footnote{based on comet dynamical type and value if its semi-major axis}&&\\
Specifics of dust particles  & compact &  porous\\
\hline
\end{tabular}
\end{minipage}
\end{table}

It is well known that a dust mantle is formed on the surfaces of cometary nuclei along with release of gas and dust near the Sun \citep[e.g.,][]{houpis-et-al85}. At the formation of the dust mantle, sublimation of volatiles in cometary nuclei may lower the porosity of dust aggregates by the packing effect, which results from anisotropic sublimation of the volatiles \citep[cf.][]{mukai-fechtig83}. Alternatively, high-porosity aggregates, such as BCCA particles, may be selectively removed from a cometary nucleus owing to their higher mobility in a gas flow compared to low-porosity aggregates, such as BPCA particles, which tend to stay in the nucleus. This produces a deficit of high-porosity aggregates in cometary nuclei, which increases as the influence of the solar radiation gets stronger and/or longer.

Because the growth of dust mantles on the surfaces of comets depends on their orbital parameters, short-period comets are likely to have low-porosity aggregates on average, while long-period comets retain high-porosity aggregates. This not only explains the fact, noticed earlier by \citet{Sitko}, that most long-period comets show silicate features in the mid-infrared spectra, while the infrared spectra of short-period comets are almost featureless; but it is also consistent with correlation between the strength of the infrared feature (as well as polarization and gas/dust ratio) and the semi-major axis for all periodic comets and similar correlation with the perihelion distance for short-period comets. This is also consistent with the strong correlation between the strength of the silicate feature and jet activity \citep{hadamcik2003, hanner2005}, since jets from active areas indicate ejection of the material from the inside, beneath the surface mantle.

Note that new comets, which we could not include in our analysis because their semi-major axes cannot be defined, show a variety of values for the strength of the silicate feature (Table \ref{tab2}). This may indicate some initial features of their parent body, which, probably, become less noticeable after several close approaches to the Sun. 

\begin{acknowledgements}
This work was supported by the DAAD grant (Kolokolova), by MPI f\"ur Sonnensystemforschung (Kiselev), and by the Ministry of Education, Culture, Science, and Technology (Monbu Kagaku Sho) (Kimura).
\end{acknowledgements}

\end{document}